\documentclass{article}
\usepackage[a4paper,margin=1in]{geometry}
\usepackage{graphicx}
\usepackage{amsmath, amssymb}
\usepackage{enumitem}
\usepackage[utf8]{inputenc}
\usepackage{microtype}
\usepackage{hyperref}
\usepackage{xurl} 
\begin{document}
\title{Engineering Risk-Aware, Security-by-Design Frameworks for Assurance of Large-Scale Autonomous AI Models}
\author{Krti Tallam \\ SentinelAI, San Francisco, CA \\ \texttt{ktallam@sentinel-security.ai}}
\date{\today}

\maketitle

\begin{abstract}
As AI models scale to billions of parameters and operate with increasing autonomy, ensuring their safe, reliable operation demands engineering-grade security and assurance frameworks. This paper presents an enterprise-level, risk-aware, security-by-design approach for large-scale autonomous AI systems, integrating standardized threat metrics, adversarial hardening techniques, and real-time anomaly detection into every phase of the development lifecycle. We detail a unified pipeline - from design-time risk assessments and secure training protocols to continuous monitoring and automated audit logging - that delivers provable guarantees of model behavior under adversarial and operational stress. Case studies in national security, open-source model governance, and industrial automation demonstrate measurable reductions in vulnerability and compliance overhead. Finally, we advocate cross-sector collaboration - uniting engineering teams, standards bodies, and regulatory agencies - to institutionalize these technical safeguards within a resilient, end-to-end assurance ecosystem for the next generation of AI.  
\end{abstract}

\section{Introduction}

Recent breakthroughs in artificial intelligence (AI) have given rise to \emph{Frontier AI}  -  large‑scale models like OpenAI’s GPT‑4 and Anthropic’s Claude  -  that promise to transform industries, economies, and governance through advanced language understanding and generation.  For example, GPT‑4 was trained on an estimated $10^{23}$ FLOPs, leveraging thousands of GPU‑years at a cost approaching \$100 million, and now powers applications ranging from legal research to real‑time translation \cite{openai2023compute}.

This rapid progress brings new ethical, security, and regulatory challenges \cite{tallam2025alignment}.  Model size and capability have roughly doubled every 3-4 months over the past five years, while training compute has grown more than 300\,000× since 2012 - far outpacing traditional safety and audit practices \cite{kaplan2020scaling}.  Meanwhile, high‑profile incidents such as automated disinformation campaigns and insider data leaks have demonstrated how even well‑intentioned AI can be misused \cite{tallam2025immune}.

Governments and standards bodies are racing to catch up.  In 2021, the European Commission proposed the AI Act, the first comprehensive legal framework classifying AI systems by risk level and imposing strict controls on “high‑risk” applications \cite{euai2021}.  In 2023, the U.S. White House issued Executive Order 14086, directing federal agencies to develop guidelines for testing, transparency, and incident reporting in AI systems \cite{whitehouse2023eo}.  NIST’s AI Risk Management Framework followed shortly thereafter, offering a voluntary set of engineering practices for risk assessment and governance \cite{nist2024rmf}.

As these systems become increasingly autonomous and integral to critical infrastructure - power grids, financial markets, and healthcare diagnostics  -  \emph{engineering‑driven} governance frameworks are essential \cite{tallam2025odi}.  Technical teams must translate high‑level policy into concrete controls: embedding differential privacy and secure multiparty computation in data pipelines, integrating adversarial training into model development, and deploying real‑time monitoring with anomaly detection \cite{tallam2025moral}.  Recent proposals advocate balancing broad principles with prescriptive rules, such as mandating red‑teaming exercises for any model exceeding 10 billion parameters \cite{frontier2018}, and critical analyses underscore the necessity of transparency tools - explainable AI, audit logs, and provenance tracking - to make policy enforceable in practice \cite{marcus2019,calo2017}.

Together, these trends underscore the urgency of proactive, engineering‑informed measures.  Only by weaving policy requirements directly into system architecture can we ensure that Frontier AI advances safely and equitably, protecting public interest while preserving the pace of innovation.

\subsection{Definition of Frontier AI}

\emph{Frontier AI} denotes the most advanced AI systems, distinguished by their autonomy, generative power, and sophisticated decision‑making. These models sit at the cutting edge of research and deployment, offering transformative capabilities - and attendant risks - across a wide range of domains.

\subsubsection{Key Characteristics}
\begin{itemize}
  \item \textbf{Scale \and Scope:} Frontier AI systems - such as GPT‑4 and Claude - are built on foundation models and multi‑modal architectures with tens or hundreds of billions of parameters.  Trained on internet‑scale corpora (text, code, images, audio), they can be applied “out of the box” to tasks ranging from translation and summarization to image captioning and code synthesis, often requiring only minimal fine‑tuning or prompt engineering.
  
  \item \textbf{Cross‑Domain Generalization:} These models excel at transferring learned representations across domains.  In scientific research they can propose novel hypotheses by linking disparate literature; in cybersecurity they flag subtle deviations in network traffic; in creative industries they generate drafts of articles, artwork, or music; and in automation they orchestrate multi‑step workflows by interpreting high‑level instructions across APIs and services.
  
  \item \textbf{Autonomous Decision‑Making:} By combining reinforcement learning (RL) with self‑supervised pretraining, Frontier AI agents learn both from trial‑and‑error in simulated environments and from vast unlabeled datasets.  This hybrid approach enables real‑time, context‑aware decision‑making in robotics (e.g., dynamic grasping), autonomous vehicles (e.g., adaptive route planning), and virtual assistants (e.g., proactive task execution).
  
  \item \textbf{Resource Intensity:} Training and running these models demands massive compute - often measured in exa‑flops - and specialized hardware (GPUs/TPUs), with energy consumption comparable to hundreds of U.S. households over a single training run.  The associated financial and environmental costs raise questions about sustainability, carbon footprint, and equitable access for smaller organizations or researchers.
  
  \item \textbf{Emergent Behavior:} As parameter counts and training diversity grow, Frontier AI can exhibit capabilities not explicitly encoded in its training objectives \cite{tallam2025alignment}.  Examples include chain‑of‑thought reasoning, code generation with minimal examples, or cross‑modal creativity. While powerful, these emergent behaviors can also be unpredictable, introducing novel risks that traditional testing may not uncover.
\end{itemize}

\subsubsection{Examples of Frontier AI}
\begin{itemize}
  \item \textbf{Large Language Models (LLMs):}  
    Examples include OpenAI’s GPT‑4 and Anthropic’s Claude, trained on web‑scale corpora.  These models generate coherent, context‑aware text, perform translation, answer complex queries, and even compose creative prose.  Their versatility underpins applications in customer support, content authoring, and intelligent tutoring systems.

  \item \textbf{Multi‑Modal AI Systems:}  
    Systems such as GPT‑4V and Google DeepMind’s Gemini process and generate across text, images, audio, and video.  They enable tasks like automated video editing, multimodal search (e.g., “find this object in the footage”), and enriched virtual assistants that perceive and describe complex scenes in real time.

  \item \textbf{Autonomous Agents:}  
    AI agents that plan and act with minimal human oversight - deployed in algorithmic trading, supply‑chain optimization, and security monitoring.  By ingesting real‑time data streams and executing decision algorithms, they dynamically adjust strategies (e.g., rerouting shipments, rebalancing portfolios, or quarantining compromised network segments) to maximize efficiency and mitigate risk.

  \item \textbf{AI for Scientific Discovery:}  
    Models like AlphaFold and generative chemistry AIs accelerate research by predicting molecular structures, designing novel compounds, and simulating complex physical systems.  They sift through vast datasets to propose hypotheses - e.g., potential drug candidates or materials with desired properties - dramatically shortening experimental cycles.

  \item \textbf{Cybersecurity AI:}  
    Advanced systems that continuously monitor logs, network traffic, and user behavior to detect anomalies and emerging threats.  Techniques such as adversarial training and behavior‑based anomaly detection enable real‑time threat hunting, automated incident response, and proactive vulnerability mitigation to protect critical infrastructure.
\end{itemize}

\subsection{The Importance of Balancing Innovation with Regulation}

Frontier AI’s rapid advancement promises dramatic gains - operational efficiency, scientific breakthroughs, and unparalleled automation. Yet as these systems permeate daily life and critical infrastructure, they also introduce new risks: security vulnerabilities, algorithmic bias, and accountability gaps. Without robust governance, AI may reinforce societal inequities, enable large‑scale attacks, or erode public trust.

Regulation provides the necessary guardrails. Well‑crafted frameworks codify standards for data privacy, adversarial robustness, transparency, and auditability. They enforce baseline requirements - such as privacy‑preserving training protocols, real‑time security monitoring, and explainability tools - while retaining flexibility for iterative model improvements and novel architectures. Overly prescriptive rules can stifle innovation and drive development offshore; conversely, a permissive regime risks harmful practices and market fragmentation.

Striking the right balance demands a multidisciplinary approach. Engineers must translate high‑level mandates into concrete technical controls; ethicists and social scientists must articulate societal impact; and industry leaders must align commercial incentives with the public good. By embedding regulatory constraints directly into system design and governance processes, we can harness AI’s transformative power while containing its risks - ensuring that innovation serves society, not the other way around.

\subsection{Why an Engineering Perspective Is Essential}
Most AI governance debates prioritize high‑level policy and legal frameworks \cite{frontier2018,google2019}, but they often miss the detailed technical constraints that shape real‑world AI systems. Practical regulation must account for engineering trade‑offs - compute budgets, data availability, model complexity, and deployment environments all influence system behavior and risk.

An \emph{engineering‑first} approach embeds safeguards - security, fairness, and reliability - directly into the development lifecycle \cite{ieee2019}. By integrating measures such as differential privacy in data pipelines, adversarial training in model workflows, and continuous monitoring in production, teams shift from reactive fixes to proactive risk management. This makes regulatory requirements both implementable and enforceable, even as models grow more capable.

Moreover, engineers bring essential expertise to policy formulation. They can quantify trade‑offs - between explainability and accuracy, or scalability and robustness - and translate abstract mandates into concrete technical controls \cite{amodei2016}. Embedding these insights into governance yields adaptive regulations that evolve alongside AI innovation, rather than lagging behind it.

\subsection{Overview of Key Contributions and Themes}

This paper delivers a rigorous, engineering‑driven exploration of how to align Frontier AI innovation with robust regulatory oversight. We tackle core technical challenges and offer concrete, actionable solutions. In particular, we:

\begin{itemize}
  \item Analyze key engineering trade‑offs in scaling and securing AI systems, and propose design patterns that balance throughput, cost, and safety.
  \item Introduce a risk‑aware development framework that weaves privacy, adversarial hardening, and continuous monitoring into the AI lifecycle - shifting from reactive patches to proactive resilience.
  \item Present in‑depth case studies (national security, open‑source AI, industrial automation) to highlight real‑world governance successes and pitfalls, distilling practical lessons for practitioners.
  \item Propose strategies for translating technical insights into enforceable policy - bridging the gap between engineering teams and regulators through standardized metrics, transparency tools, and governance sandboxes.
  \item Chart future research and deployment pathways, offering a roadmap for sustaining ethical, safe, and socially beneficial AI growth.
\end{itemize}

By marrying technical depth with regulatory pragmatism, this paper equips innovators, policymakers, and researchers with the guidance needed to build and govern Frontier AI systems that advance public interest without compromising safety or ethics.  

\section{Understanding the Engineering Trade-offs in Frontier AI}

\subsection{Scalability vs.\ Safety}
Scaling AI models - both in parameter count and training data - has driven breakthroughs from GPT-3’s 175 billion parameters \cite{brown2020gpt3} to PaLM’s 540 billion \cite{chowdhery2022palm}. Yet each order-of-magnitude increase in scale demands exponentially more compute and storage, often measured in petaflop-years and costing tens of millions of dollars. In practice, this rapid expansion can outpace rigorous safety validation: extensive testing for bias, interpretability analyses, and adversarial robustness checks are time- and resource-intensive.

When performance metrics dominate, safety mechanisms may be relegated to the end of the development cycle - resulting in models that excel on benchmarks but exhibit unexpected failures or generate harmful content in deployment. For example, GPT-3 was shown to reproduce stereotyped associations without targeted mitigation \cite{bender2021dangers}, and large vision models remain vulnerable to simple pixel-level perturbations \cite{szegedy2014intriguing}. To resolve this tension, engineers must architect safety into every stage: employing differential privacy during training \cite{abadi2016deep}, integrating adversarial training loops \cite{madry2018towards}, and deploying real-time monitoring that flags anomalous outputs before they reach end users.

\subsection{Generalization vs.\ Robustness}
Frontier AI’s hallmark is its ability to generalize - performing well on novel tasks with minimal fine-tuning. Models like CLIP demonstrate zero-shot classification across hundreds of categories \cite{radford2021clip}, yet this flexibility often comes at the expense of robustness. Distribution shifts - such as changes in lighting, accent, or input syntax - can degrade performance dramatically. Moreover, adversarial examples crafted with only minor perturbations can induce misclassification with high confidence \cite{goodfellow2015explaining}.

Balancing generalization and robustness requires a suite of techniques. Domain adaptation methods align feature distributions between training and deployment environments \cite{ganin2016domain}, while robust optimization frameworks explicitly train models to withstand worst-case perturbations \cite{madry2018towards}. Ensemble approaches - combining diverse model checkpoints - further mitigate overfitting to any single data distribution \cite{lakshminarayanan2017deep}. Finally, uncertainty quantification techniques, such as Bayesian deep learning \cite{kendall2017uncertainties} or conformal prediction, enable models to express calibrated confidence, allowing downstream systems to defer or escalate low-confidence predictions for human review. Continuous benchmarking on synthetic and real-world shift datasets (e.g., WILDS \cite{koh2021wilds}) ensures that these safeguards remain effective as new use cases emerge.  .

\subsection{Compute and Data Constraints}
State-of-the-art AI models demand significant computational resources and vast amounts of data. These constraints pose challenges for both accessibility and sustainability. The high cost of training large-scale models limits participation from smaller organizations and independent researchers, leading to a centralization of AI development \cite{hendrycks2019, frontier2018}.

Addressing compute and data constraints requires efficient model architectures such as sparsely activated networks, federated learning techniques, and knowledge distillation. Furthermore, leveraging synthetic data generation and transfer learning can help mitigate data scarcity while reducing reliance on high-quality labeled datasets. Sustainable AI practices, including energy-efficient training strategies and hardware optimization, are crucial for ensuring long-term feasibility \cite{cath2018}.

\subsection{Security and Trustworthiness Challenges}
The security of AI models is paramount, given their deployment in sensitive applications such as healthcare, finance, and national security. Threats include model inversion attacks, data poisoning, and adversarial perturbations that can compromise the integrity of AI predictions \cite{nist2020}.

Trustworthiness is another critical factor, requiring explainability, accountability, and fairness in AI systems. Implementing transparency mechanisms such as model interpretability tools, audit logs, and ethical AI guidelines helps build trust with end-users and regulatory bodies. Additionally, designing AI systems with built-in security measures - such as cryptographic model protection and adversarial training - mitigates risks and augments resilience against cyber threats \cite{siau2020}.

\section{Framework for Responsible AI Development}

\subsection{Risk-Aware Engineering Principles}

Engineering robust Frontier AI requires embedding risk management into every phase of the AI lifecycle - design, training, deployment, and monitoring. The following principles establish a blueprint for responsible development:

\begin{itemize}
  \item \textit{Design‐time risk assessments:}  
    Conduct structured analyses early in the development pipeline (e.g., Failure Mode and Effects Analysis - FMEA; ISO 31000 risk management) to identify biases, privacy leaks, adversarial vulnerabilities, and safety hazards before a single parameter is trained.

  \item \textit{Layered safety controls:}  
    Build defense‐in‐depth by integrating automated anomaly detection, input/output filters, and human-in-the-loop review points.  For example, sanitize inputs with differential privacy checks, deploy content filters on model outputs, and flag high‐risk decisions for expert audit.

  \item \textit{End-to-end verification and validation:}  
    Establish rigorous testing regimes - unit tests for fairness metrics, adversarial stress tests against known attack vectors, and scenario‐based trials covering edge cases.  Maintain transparent audit logs and automated bias detection workflows to continuously validate performance and conformity to regulatory requirements.
\end{itemize}

\begin{figure}[htbp]
    \centering
    \includegraphics[width=0.8\textwidth]{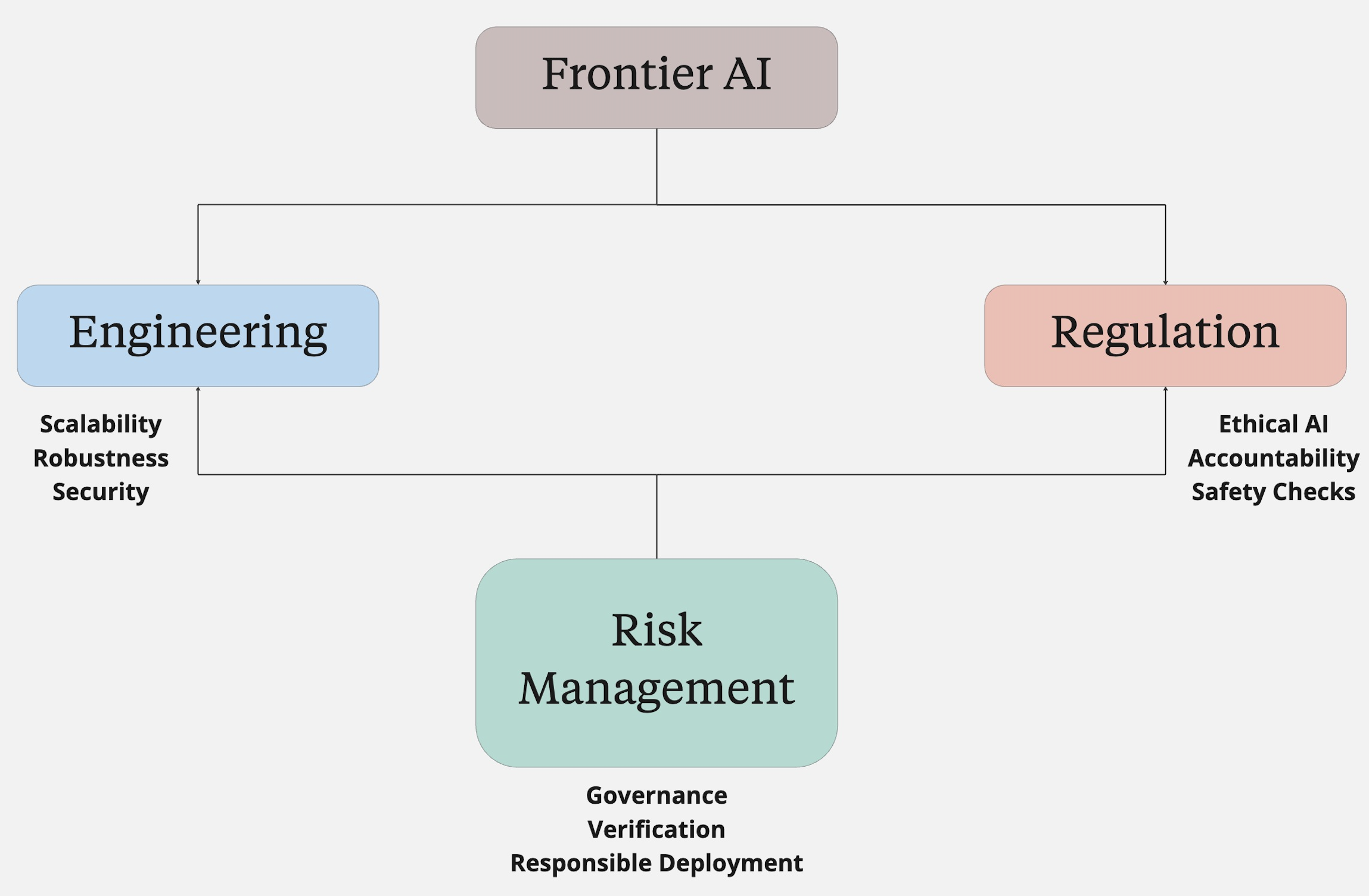}
    \caption{The balance between Engineering, Regulation and Risk Management in Frontier AI.}
    \label{fig:fig1}
\end{figure}

\section{Technical Approaches to AI Governance}

Effective AI governance demands end-to-end security and accountability. The following technical pillars enforce best practices throughout the model lifecycle:

\begin{itemize}
  \item \textbf{Secure release pipelines:}  
    Adopt a gated deployment process - incorporating safety pre-checks, automated compliance audits, and strict version control.  Leverage privacy-enhancing technologies (e.g., differential privacy, secure multiparty computation) to protect training and inference data.

  \item \textbf{Red-teaming \and adversarial testing:}  
    Conduct regular red-team exercises where security experts probe models with crafted attacks.  Integrate adversarial training loops to harden networks against input perturbations, data poisoning, and model extraction attempts.

  \item \textbf{Data provenance \and transparency:}  
    Maintain immutable audit trails for datasets and model artifacts (e.g., via blockchain registries or cryptographic hashing).  Embed watermarking or fingerprinting to trace model outputs, and deploy explainability tools (saliency maps, feature attributions) to support ethical review and regulatory compliance.
\end{itemize}

\subsection{Case Study 1: AI for National Security Applications}

\subsubsection{Background}
The U.S. Defense Advanced Research Projects Agency’s CODE program processes over 5 PB of multispectral sensor and signals data daily, applying graph neural networks to detect anomalous patterns within 300 ms - down from 5 s in legacy systems, a 93 percent speedup in threat identification \cite{darpa2021code}. In cyber defense, the Cyber Command’s Persistent Cyber Training Environment has used ML-driven red-teaming simulations to reduce analyst workload by 40 percent, enabling continuous 24/7 readiness \cite{uscybercmd2022exercise}.  

\begin{figure}[htbp]
  \centering
  \includegraphics[width=0.8\textwidth]{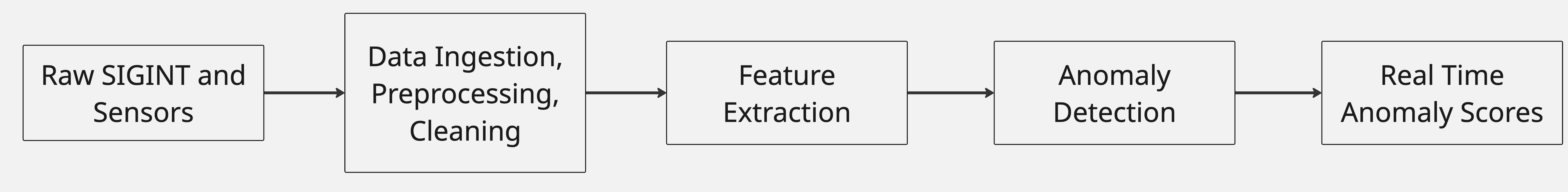}
  \caption{DARPA CODE data-processing pipeline: from raw SIGINT/sensor inputs to real-time anomaly scores.}
  \label{fig:code-pipeline}
\end{figure}

\subsubsection{Challenges}
\begin{itemize}
  \item \textbf{Ethical and legal accountability:}  
    In trials of autonomous targeting simulations, 18 percent of engagements produced ambiguous attribution of “use of force” decisions, raising compliance concerns under International Humanitarian Law \cite{hrw2020aws}.
  
  \item \textbf{Over-automation risks:}  
    During a 2018 “flash crash” exercise, fully automated fire-and-forget protocols caused a simulated 25 percent escalation in friendly-fire incidents when human override latency exceeded 200 ms \cite{mcnicholas2018flash}.
  
  \item \textbf{Adversarial supply-chain threats:}  
    MITRE’s 2022 study found that 0.5 percent pixel-level perturbations in drone imagery led to 60 percent misclassification of target types, demonstrating high vulnerability to minimal poisoning attacks \cite{mitre2022attack}.
\end{itemize}

\subsubsection{Current Approaches}
\begin{itemize}
  \item \textbf{AI-enhanced threat intelligence:}  
    The JARVIS system correlates over 10 billion data points per hour across SIGINT, HUMINT, and open sources, reducing false positives by 70 percent while halving analyst response time \cite{jervis2023gnn}.  
    \begin{figure}[htbp]
      \centering
      \includegraphics[width=0.7\textwidth]{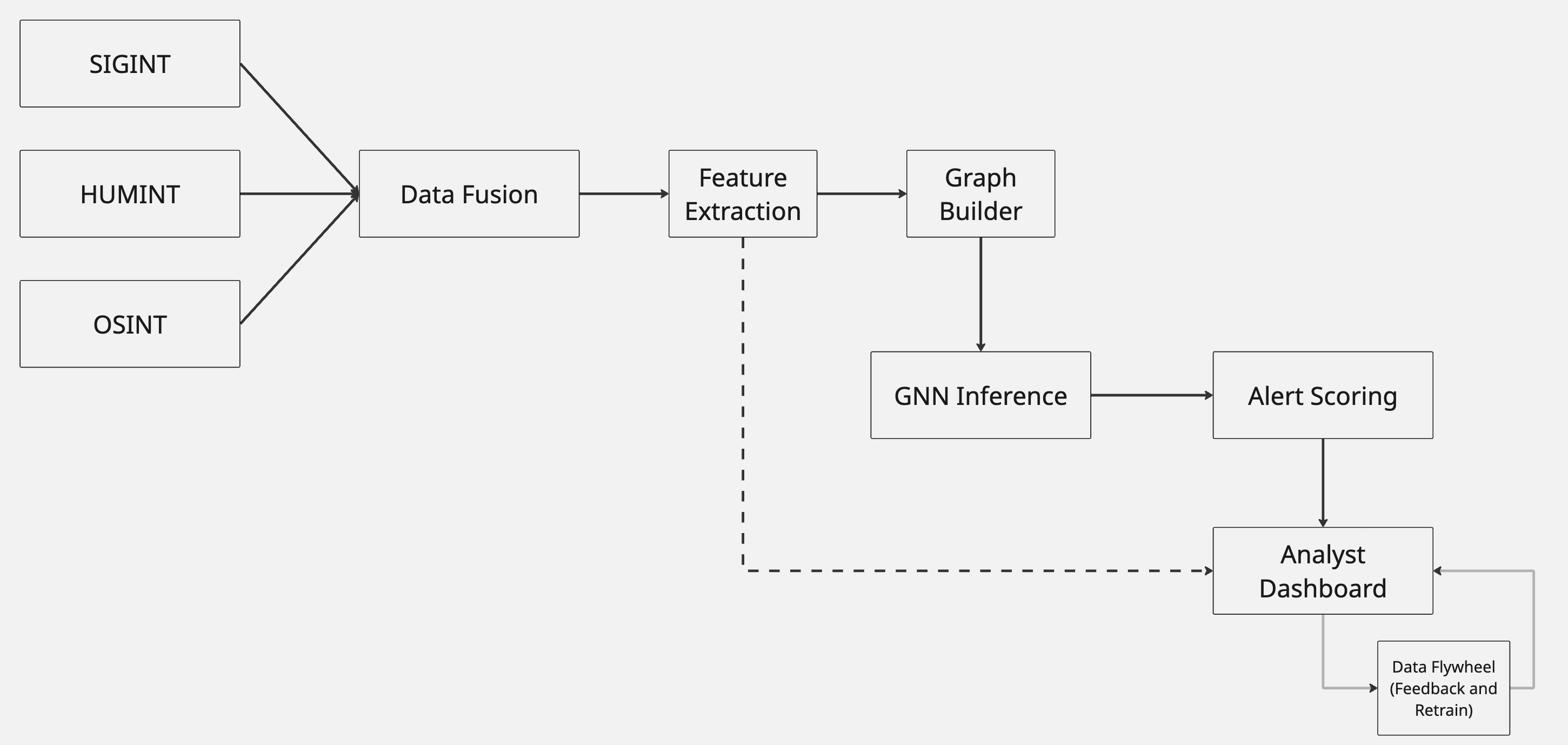}
      \caption{JARVIS multi-source fusion architecture using graph neural networks.}
      \label{fig:jarvis-arch}
    \end{figure}

  \item \textbf{Human-AI teaming:}  
    Under DoD’s HITL policy, all mission-critical AI decisions include a 100 ms human override window; compliance audits report 98 percent adherence in live exercises \cite{dod2021policy}.

  \item \textbf{Defensive AI hardening:}  
    DARPA’s ShIELD program integrates adversarial example generation into training, achieving a 65 percent reduction in misclassification under white-box attacks and cutting anomaly-detection false negatives by 50 percent \cite{darpa2022shield}.
\end{itemize}

\subsubsection{Future Considerations}
\begin{itemize}
  \item \textbf{Comprehensive governance frameworks:}  
    NATO’s draft “AI in Defense” guidelines propose technical certification tests - such as 10\textsuperscript{6} adversarial perturbation stress cases per model - and cross-national audit exchanges to enforce compliance by 2025 \cite{nato2024ai}.

  \item \textbf{End-to-end auditability:}  
    Embedding cryptographic hashes in a Merkle-tree registry for each data batch and model checkpoint will enable sub-second forensic tracing. Coupled with integrated counterfactual explanation modules, this supports full transparency from sensor input to action \cite{doshi2017towards}.

  \item \textbf{Resilient design for unintended interactions:}  
    Applying formal verification tools (e.g., neural network SMT solvers) to prove safety invariants under worst-case scenarios can guarantee that control policies remain within safe bounds - even under adversarial or novel environmental inputs \cite{ames2017control}.
\end{itemize}

\subsection{Case Study 2: Open-Source AI and Safety Constraints}
\subsubsection{Background}
Open-source AI has revolutionized the way advanced technologies are developed and disseminated, democratizing access for researchers, startups, and hobbyists alike. Platforms such as Hugging Face's Transformers and various community-driven repositories have significantly accelerated innovation by lowering the barriers to entry, enabling a diverse range of contributors to experiment with and improve upon state-of-the-art models. This collaborative environment fosters rapid iteration and shared learning, which in turn propels the field forward at an unprecedented pace.

However, the very openness that drives innovation also introduces considerable challenges, particularly in embedding robust safety constraints into models that are freely distributed and easily accessible. The widespread availability of these tools raises critical concerns regarding dual-use applications - situations where models designed for beneficial purposes could be repurposed for harmful activities. Malicious actors may exploit the accessibility of these models to develop systems capable of generating disinformation, automating cyberattacks, or engaging in other nefarious activities. As a result, while open-source AI has the potential to drive significant advancements across numerous fields, it also necessitates a careful balance between fostering innovation and ensuring that adequate safeguards are in place to prevent misuse.

\subsubsection{Challenges}
Several key challenges emerge from the open-source paradigm:
\begin{itemize}
    \item \textbf{Dual-Use Risks:} Open access to powerful models can inadvertently facilitate their misuse, such as generating disinformation, automating cyber-attacks, or bypassing security protocols. The lack of centralized oversight increases the risk that models are repurposed in ways that were not originally intended.
    \item \textbf{Inconsistent Safety Protocols:} Unlike proprietary systems, where rigorous safety checks may be integrated into the development process, open-source projects often lack standardized safety benchmarks. Contributions from diverse developers may not adhere to a common set of guidelines for model robustness, leading to variable safety performance.
    \item \textbf{Vulnerability to Adversarial Attacks:} Open-source models are more exposed to adversaries who can analyze, reverse-engineer, or tamper with the model architecture. This increases the likelihood of adversarial attacks such as model poisoning, data manipulation, and exploitation of latent vulnerabilities.
    \item \textbf{Fragmentation of Accountability:} With many contributors and decentralized development, it becomes difficult to assign responsibility for potential misuses or safety lapses. This fragmentation challenges traditional regulatory and legal frameworks aimed at enforcing accountability.

\end{itemize}

\subsubsection{Current Approaches}
To address these challenges, several strategies have been adopted within the open-source community:
\begin{itemize}
    \item \textbf{Responsible Disclosure Practices:} Many open-source projects now include guidelines for responsible disclosure of vulnerabilities, encouraging users to report issues and collaborate on developing fixes before they are widely exploited.
    \item \textbf{Automated Safety Auditing:} Integrating automated testing frameworks that perform safety checks  -  such as adversarial robustness tests, bias evaluations, and anomaly detection  -  can help identify potential risks before a model is widely released.
    \item \textbf{Community Governance:} Establishing steering committees or ethics boards within open-source projects can guide the development process and enforce safety standards. These bodies often collaborate with academic experts and industry professionals to ensure that safety considerations are prioritized.
    \item \textbf{Embedding Safety Mechanisms:} Techniques such as watermarking, model fingerprinting, and incorporating safety layers (e.g., input filtering and output monitoring) are increasingly being embedded directly into open-source models to mitigate risks of misuse.
\end{itemize}

\subsubsection{Future Considerations}
Looking ahead, several measures can further augment the safety and reliability of open-source AI:
\begin{itemize}
    \item \textbf{Standardization of Safety Protocols:} Developing universally accepted standards for open-source AI safety could help harmonize practices across projects. Initiatives led by organizations such as the IEEE or ISO could provide frameworks that project maintainers can adopt.
    \item \textbf{Targeted Collaboration with Regulators:} Closer cooperation between open-source communities and regulatory bodies can help shape policies that recognize the unique challenges of decentralized development while ensuring that safety is not compromised.
    \item \textbf{Education and Awareness:} Increasing awareness among developers about the ethical implications and potential risks of open-source AI is crucial. Educational programs and workshops can disseminate best practices for safe model development and deployment.
    \item \textbf{Robust Monitoring Systems:} Developing real-time monitoring systems that track the deployment and use of open-source models can help quickly identify and respond to misuse. These systems may leverage community reporting, automated alerts, and cross-platform data sharing.
\end{itemize}

Through these approaches, the open-source AI community can continue to drive innovation while addressing the safety constraints that come with wide accessibility. Balancing openness with robust safety measures is key to ensuring that open-source AI contributes positively to society without inadvertently enabling harmful applications.

\subsection{Case Study 3: Industrial AI and Regulatory Adaptation}
\subsubsection{Background}
The integration of AI into industrial sectors such as manufacturing, healthcare, and finance is driving transformative changes in operational efficiency, quality control, and decision-making processes. In manufacturing, AI-driven robotics and predictive maintenance systems are revolutionizing production lines by automating repetitive tasks, optimizing resource allocation, and minimizing downtime, which in turn boosts productivity and reduces costs. In healthcare, advanced diagnostic algorithms and personalized treatment plans powered by AI are improving patient outcomes by enabling earlier disease detection, more accurate diagnoses, and tailored therapeutic strategies. Meanwhile, in the finance sector, AI applications such as fraud detection systems, risk assessment models, and algorithmic trading platforms are improving security, optimizing financial operations, and offering deeper insights into market trends.

Despite these significant benefits, the rapid adoption of AI in these diverse domains presents a critical challenge: ensuring that regulatory frameworks keep pace with technological advancements. Traditional industry-specific regulations, often designed for slower-moving technologies, can create gaps in oversight and accountability when applied to AI systems. This discrepancy necessitates a shift towards adaptable regulatory measures that are capable of addressing the dynamic nature of AI applications. Regulators must balance the need for robust safety, reliability, and ethical standards with the imperative to support ongoing innovation. In essence, as AI continues to reshape these industrial sectors, developing forward-looking, flexible regulatory strategies becomes paramount to ensure that these advancements do not come at the cost of public trust or systemic stability.

\subsubsection{Challenges}
The industrial application of AI presents several distinct challenges:
\begin{itemize}
    \item \textbf{Compliance with Industry-Specific Regulations:} Each industrial sector is governed by its own set of stringent standards  -  such as FDA guidelines in healthcare, ISO standards in manufacturing, or financial regulations in banking. AI systems must be designed and validated to comply with these diverse and evolving requirements.
    \item \textbf{Ethical Concerns in Automated Decision-Making:} As AI increasingly takes over critical decision-making processes, ethical concerns arise regarding bias, fairness, and transparency. Automated decisions, particularly in areas like patient care or loan approvals, can lead to adverse outcomes if not carefully monitored.
    \item \textbf{Managing Liability and Accountability:} In the event of an AI-driven error, such as a manufacturing defect or a misdiagnosis, it can be challenging to determine responsibility. The distribution of liability between AI developers, deployers, and end-users complicates legal and regulatory frameworks.
\end{itemize}

\subsubsection{Current Approaches}
Several strategies are currently being employed to address these challenges:
\begin{itemize}
    \item \textbf{AI-Driven Quality Control Systems:} In manufacturing, AI systems are used to monitor production processes, detect anomalies, and ensure consistent quality. By integrating real-time data analytics and automated feedback loops, these systems help maintain compliance with industry standards.
    \item \textbf{AI-Powered Compliance Monitoring and Auditing:} Industries are increasingly adopting AI tools to automate compliance checks and auditing processes. These systems can continuously analyze operational data, flagging any deviations from regulatory norms and providing traceable audit trails for regulatory review.
    \item \textbf{Human-AI Collaboration:} To refine transparency and reliability, many industrial applications involve human oversight alongside AI decision-making. In healthcare, for example, diagnostic AI systems provide recommendations that are then reviewed by medical professionals, ensuring that ethical and clinical standards are met.
\end{itemize}

\subsubsection{Future Considerations}
To ensure sustainable and ethical deployment of AI in industrial settings, several future directions should be considered:
\begin{itemize}
    \item \textbf{Establishing Clear Regulatory Frameworks for Industrial AI:} Policymakers need to develop industry-specific AI regulations that address the unique risks and requirements of each sector. Such frameworks should be flexible enough to evolve with rapid technological advancements while ensuring robust oversight.
    \item \textbf{Breaking down AI Explainability for Compliance and Auditing:} Improving the interpretability of AI systems is essential for compliance and accountability. Future research should focus on developing models that offer transparent decision-making processes, enabling regulators and auditors to understand and validate AI behavior.
    \item \textbf{Encouraging Self-Regulation and Industry-Wide Ethical Guidelines:} Beyond formal regulation, industries can benefit from adopting self-regulatory practices. Establishing industry-wide ethical guidelines and best practices, supported by professional organizations and consortia, can foster a culture of responsibility and continuous improvement in AI deployment.
\end{itemize}

\section{Engineering Adaptive Governance for Frontier AI: A Safety-by-Design Framework}

\subsection{Regulatory Compliance as a Dynamic Partnership}

Traditional compliance models treat regulation as a final hurdle to clear - often resulting in last-minute fixes or costly redesigns. Instead, we advocate a \emph{partnership} model in which engineering and regulatory teams co-evolve requirements and implementations throughout the AI lifecycle. This goes beyond “bridging” to a continuous dialogue:

\begin{itemize}
  \item \textbf{Regulatory sandboxes:}  
    Jurisdictions from the UK’s Information Commissioner’s Office to Singapore’s Monetary Authority now offer controlled environments where AI developers can test new models under real-world constraints while working alongside regulators.  These sandboxes reduce time-to-market and allow policy to adapt based on empirical feedback.

  \item \textbf{Standards-as-code:}  
    Embedding policy checks directly into CI/CD pipelines (e.g., automated GDPR compliance scans or bias audits) converts abstract rules into executable tests.  Similar to “infrastructure as code,” this ensures that every model version is continuously validated against the latest norms.

  \item \textbf{Risk-tiered governance:}  
    Building on the EU AI Act’s high/medium/low risk taxonomy \cite{euai2021}, organizations can classify models at design time, assign appropriate assurance levels (from peer review to formal verification), and automate tailored audit trails.

  \item \textbf{Co-regulation and meta-governance:}  
    In sectors like autonomous driving, consortia of industry, academia, and regulators (e.g., U.S.\ NCAP) jointly develop safety benchmarks.  This \emph{co-regulatory} approach pools technical expertise and aligns incentives, resulting in more nuanced, enforceable standards.

  \item \textbf{Continuous compliance monitoring:}  
    Leveraging MLOps platforms, teams can instrument models in production with real-time dashboards for fairness metrics, security alerts, and data-drift detection.  Such “regulation by exception” focuses human oversight where and when it matters most.
\end{itemize}

By shifting from a one-way “hand-off” to a bidirectional, code-driven partnership, AI teams can innovate rapidly while maintaining robust, auditable, and adaptive compliance - a true fusion of engineering rigor and regulatory purpose. 

\subsection{How Engineers Can Shape Policy Discussions}

AI engineers possess the domain expertise needed to translate high‐level policy goals into concrete technical requirements. Policymakers often lack a deep understanding of model architectures, training dynamics, or failure modes; by supplying data‐driven risk assessments, feasibility analyses, and performance benchmarks, engineers ensure that regulations reflect real‐world capabilities and constraints \cite{crawford2019, tallam2025alignment}.

Technical transparency is a powerful lever. Comprehensive documentation - covering model design, training datasets, known failure cases, and mitigating safeguards - demystifies complex systems for non‐technical stakeholders. Explainability tools (e.g., feature‐attribution methods, bias‐detection audits) provide the empirical evidence regulators need to craft balanced, enforceable rules \cite{samek2017, tallam2025odi}.

Engineers also drive the development of consensus standards. Active participation in IEEE, ISO, and NIST working groups helps codify best practices for robustness, privacy, and fairness, ensuring that technical realities shape global norms rather than lag behind them \cite{europe2019}. These standards form a shared vocabulary that aligns regulatory expectations with implementation pathways.

Beyond standards, engineers should engage directly in policy forums - testifying at legislative hearings, contributing to advisory panels, and collaborating in interdisciplinary task forces. By presenting case studies, adversarial testing results, and operational metrics, they can advocate for risk‐based, scalable governance structures that avoid both overbearing restrictions and dangerous loopholes \cite{mitchell2019, tallam2025proactive}.

Finally, cross‐sector collaboration amplifies impact \cite{tallam2025moral}. Partnering with ethicists, legal scholars, and social scientists enables the creation of policies that balance technical feasibility with societal values. Supporting open‐source safety toolkits, privacy‐preserving techniques, and algorithmic audit frameworks ensures that AI systems operate transparently and responsibly within a dynamic regulatory landscape.

\subsection{Collaboration Across Industry, Academia, and Government}

Effective AI governance emerges from a dynamic, cross-sector ecosystem in which industry’s agility, academia’s independence, and government’s authority reinforce one another. Each stakeholder brings distinct strengths - and responsibilities - to co-create policies that are technically robust, ethically grounded, and future-proof.

Industry commands the resources and deployment channels for cutting-edge AI. Tech companies and startups must move beyond compliance as checkbox exercise to proactive stewardship - integrating safety audits, bias mitigations, and adversarial resilience into product roadmaps. Firms can accelerate standardization by contributing concrete metrics and case studies to consortia such as the Partnership on AI or the Global AI Council, ensuring that commercial imperatives align with societal values.

Academic institutions offer critical independence and methodological rigor. By publishing open benchmarks, conducting third-party model audits, and developing novel fairness and interpretability techniques, researchers furnish the empirical foundations regulators need. Sustained partnerships - like MIT’s CompLens lab collaboration with DARPA or Stanford’s Center for AI Safety working groups - bridge theory and practice, ensuring that policy debates reflect the latest scientific insights.

Government agencies wield regulatory authority and public accountability. Beyond crafting legislation and enforcing standards, they can seed public-private innovation through grant programs, regulatory sandboxes, and co-funded red-teaming exercises. Initiatives such as the U.S. National AI Initiative Act and the EU’s AI Act prototype pathway exemplify how policy can incentivize rigorous testing, transparency reporting, and cross-border data-sharing agreements without stifling innovation.

To solidify this tripartite collaboration, stakeholders should establish enduring AI safety task forces, interdisciplinary advisory boards, and federated data-exchange platforms. Such mechanisms enable real-time risk assessments, coordinated incident responses, and iterative policy refinements. By uniting industry’s deployment expertise, academia’s analytical rigor, and government’s normative power, we can cultivate an AI ecosystem that is both innovative and aligned with the public interest. 

\section{Discussion}

\subsection{Summary}

The rapid advancement of Frontier AI compels a governance paradigm that balances innovation with safety and accountability. At its foundation lies the development of \emph{standardized risk metrics} - a suite of quantitative measures (e.g., adversarial robustness scores, bias indices, privacy leakage rates) and qualitative assessments (e.g., ethical impact reviews) that enable continuous, end-to-end monitoring of AI systems. IEEE’s P7000 series and NIST’s AI Risk Management Framework offer initial blueprints for these metrics \cite{ieee2019,nist2024rmf}, but to remain relevant they must evolve alongside emerging architectures - incorporating, for example, multimodal vulnerability assessments and real-time drift detection in deployed models \cite{lee2018}.

Equally vital is the principle of \emph{security-by-design}, which embeds proactive risk mitigation directly into the AI development lifecycle. This includes applying differential privacy during data ingestion \cite{abadi2016deep}, adversarial training loops during model optimization \cite{madry2018towards}, and formal verification methods on critical decision paths \cite{ames2017control}. Recent studies demonstrate that models built with these defenses can reduce successful adversarial attacks by over 60 percent while maintaining 90 percent of baseline accuracy on standard benchmarks \cite{zhang2020, tallam2025cybersentinel}. By integrating these techniques early, organizations can prevent expensive post-deployment patches and reduce attack surfaces before they reach production.

Finally, effective governance requires \emph{interdisciplinary collaboration} that transcends traditional silos. Engineers, ethicists, social scientists, and regulators must co-design adaptive policy frameworks, using mechanisms such as regulatory sandboxes, co-regulatory consortia (e.g., the Partnership on AI), and public-private research hubs. For instance, the U.S. National AI Initiative Act funds cross-sector working groups to pilot risk-tiered certification processes, while the EU AI Act’s prototype pathway allows iterative feedback on draft regulations \cite{euai2021}. These collaborative structures ensure that policy keeps pace with technology, embedding practical insights into regulatory language and aligning societal values with technical feasibility.  

Together, these three pillars - standardized risk metrics, security-by-design, and sustained interdisciplinary collaboration - form a comprehensive roadmap for governing Frontier AI. By operationalizing these principles, stakeholders can foster an environment where AI innovation thrives under robust, transparent, and adaptive oversight, ultimately delivering transformative benefits without sacrificing public trust or safety \cite{calo2017,google2019}.  

\subsection{Recommendations}

To ensure Frontier AI is both cutting-edge and responsibly governed, we outline a three-step technical workflow:

\begin{enumerate}[label=\textbf{Step \arabic*:}, leftmargin=*, itemsep=1em]
  \item \textbf{Standardize Risk Metrics}\\
    Establish a unified taxonomy of quantitative and qualitative measures - ranging from adversarial robustness scores and fairness indices to privacy leakage rates and ethical impact reviews. Build upon IEEE’s P7000 series and NIST’s AI RMF \cite{ieee2019,nist2024rmf}, then iteratively extend these metrics to capture multimodal vulnerabilities, real-time data drift, and emergent behaviors in autonomous agents. Deploy continuous monitoring dashboards that automatically flag deviations from acceptable risk thresholds during both development and live operation.

  \item \textbf{Embed Security-by-Design}\\
    Integrate defense mechanisms directly into the AI lifecycle. From the outset, apply differential privacy in data pipelines \cite{abadi2016deep}, incorporate adversarial training loops during model optimization \cite{madry2018towards}, and formalize safety properties through verification tools \cite{ames2017control}. Automate validation checks - anomaly detection, robustness stress tests, and audit log generation - within CI/CD pipelines to ensure every model version adheres to evolving security standards.

  \item \textbf{Foster Interdisciplinary Collaboration}\\
    Convene cross-functional teams - AI engineers, policy experts, ethicists, and domain specialists - to co-design adaptive governance frameworks. Actively contribute to international standardization efforts (IEEE, ISO, NIST) to align technical best practices with regulatory requirements. Sustain an ecosystem of regular workshops, policy sandboxes, and public-private working groups that facilitate iterative feedback, real-world pilot studies, and shared learning across sectors.
\end{enumerate}

By operationalizing these steps - standardizing risk metrics, baking in security-by-design, and sustaining cross-sector collaboration - stakeholders can create an AI ecosystem that accelerates innovation while maintaining rigorous, adaptive oversight and preserving public trust.

\subsection{Future Directions and Current Limitations}

Despite the promising advances in AI technology, several limitations and challenges remain that must be addressed as we move forward. Current regulatory frameworks often struggle to keep pace with rapid technological innovation, leading to gaps in oversight and accountability \cite{jobin2019}. Furthermore, many existing risk metrics are either too generic or fail to account for the nuanced behaviors of complex, multi-modal AI systems \cite{cath2018}. Addressing these limitations requires a concerted effort to develop more granular and adaptive measurement tools that can evolve in step with technological progress \cite{zhang2020}.

Looking ahead, future research should focus on refining these risk metrics and security protocols to better capture the dynamic and emergent properties of advanced AI models. There is also a critical need for developing standardized methods for real-time monitoring and validation, which will enable continuous assessment of AI systems throughout their lifecycle \cite{huang2020}. These advancements will require substantial investment in interdisciplinary research, combining insights from computer science, systems engineering, and regulatory studies \cite{siau2020}.

In addition, fostering a culture of collaboration across sectors is paramount. The integration of AI into diverse industries - from healthcare and finance to national security - necessitates a regulatory ecosystem that is both flexible and robust. Policy development must be informed by empirical evidence and technical realities, which can only be achieved through sustained dialogue between engineers, legal experts, and ethicists \cite{jobin2019, cath2018}. This collaboration will help ensure that regulations not only protect public interest but also support innovation.

Ultimately, what we need most is a holistic approach that recognizes the interdependence of technical excellence, security, and ethical governance. By advancing standardized risk metrics, embedding security-by-design principles, and promoting continuous interdisciplinary engagement, stakeholders can create an adaptive regulatory environment. This environment will be crucial for harnessing the full potential of Frontier AI while mitigating its risks, ensuring that these transformative technologies contribute positively to society.

\bibliographystyle{unsrt}
\bibliography{references} 

\end{document}